\documentclass[aps,pra,showpacs,twocolumn]{revtex4-1}
\usepackage{amssymb}
\usepackage{amsmath}
\usepackage{graphicx}
\usepackage{epsfig}

\begin{document}

\title{Amplitude control of a quantum state in non-Hermitian Rice-Mele model
driven by an external field}
\author{S. Lin${}^1$, X. Z. Zhang,${}^2$ and Z. Song${}^1$}
\email{songtc@nankai.edu.cn}
\affiliation{${\ }^1$School of Physics, Nankai University, Tianjin 300071, China \\
${\ }^2$College of Physics and Materials Science, Tianjin Normal University, Tianjin 300387, China}

\begin{abstract}
In the Hermitian regime, the Berry phase is always a real number. It may be
imaginary for a non-Hermitian system, which leads to amplitude amplification
or attenuation of an evolved quantum state. We study the dynamics of the
non-Hermitian Rice-Mele model driven by a time-dependent external field. The
exact results show that it can have full real spectrum for any value of the
field. Several rigorous results are presented for the Berry phase with
respect to the varying field. We demonstrate that the Berry phase is the
same complex constant for any initial state in a single sub-band. Numerical
simulation indicates that the amplitude control of a state can be
accomplished by a quasi-adiabatic process within a short time.
\end{abstract}

\maketitle

%%%{11.30.Er Charge conjugation, parity, time reversal, and other discrete symmetries}
%03.65.vf Phases: geometric; dynamic or topological
%03.65.-w Quantum mechanics

\section{Introduction}

In the Hermitian regime, the Berry phase \cite{Pancharatnam,Berry,Xiao} is
always the real number. It may be complex for a non-Hermitian system, the
concept of which was first introduced by Garrison and Wright \cite{Garrison}
in a dissipative system. And in these years the geometric phase for quantum
systems governed by non-Hermitian Hamiltonians and complex-valued geometric
phase effects have gained considerable attention and have been studied by
various authors \cite%
{Dattoli,Ning,Moore,Gao,Pont,Massar,Ge,Whitney,Mehri-Dehnavi,Nesterov,Cui,Liang,LC}%
. The existence of imagine part of the phase could lead to amplitude
amplification or attenuation of an evolved quantum state. We investigate the
dynamics of the non-Hermitian Rice-Mele model driven by a time-dependent
external field. The exact results show that it can have full real spectrum
for any value of the field and several rigorous results are presented for
the Berry phase with respect to the varying field. We find that the Berry
phase is the same complex constant for any initial state in a single
sub-band. And via numerical simulations, the amplitude control of a state
can be achieved by a quasi-adiabatic process within a short time.

In a Hermitian quantum system, the geometric phase acquired during an
adiabatic evolution is always real and can bring nothing to an evolved
quantum state if only one eigenstate is involved. It has been shown that a
non-Hermitian system can have real spectrum \cite{Bender1} and possess
peculiar phenomena. These include fast propagation \cite{Bender2}, infinite
reflection coefficient \cite{Ali, Longhi1, ZXZ}, unidirectional transmission
\cite{LXQ}, transmission phase lapse \cite{ZG}, maximum multi-particle
entanglement associated with the phase transition \cite{Tony} as well as the
complex Berry phase. The imaginary part of the Berry phase is significant
for a propagating particle since it may be utilized to directly amplify or
attenuate the particle probability. Very recently, the spectral and
dynamical properties of a quantum particle constrained on a ring threaded by
a time-varying magnetic flux in the presence of a complex (non-Hermitian)
potential are investigated \cite{Longhi2}. It has been demonstrated that
several striking effects are observed in the non-Hermitian case in
comparison with the Hermitian one.

In a previous work \cite{HWH1}, the dynamical behavior has been investigated
for a non-Hermitian Rice-Mele model in the absence of an external magnetic
field. It has been shown that within the unbroken $\mathcal{PT}$-symmetric
region, the translational symmetry ensures the probability-preserving
evolution of a state, which involves only one sub-band or two sub-bands with
different $k$. In this paper we aim at investigating the dynamical behaviors
in the same model but in the presence of a time-varying flux. We have
determined that the law of probability preservation still holds in the
presence of a constant flux. When the flux changes adiabatically, we will
show rigorously that the Berry phases of all the eigenstates within a
sub-band are identical complex numbers, which depend on the combination of
the system parameters, including the lattice distortion, imaginary
potential, and flux. The imaginary Berry phase leads to the amplification
and attenuation of the amplitude of an evolved quantum state. In contrast to
the review \cite{Rotter} considered in the situation where the Hamiltonian
has singularities, i.e., diabolic or exceptional points, and all the Berry
phases in the review are encircling these special points, we would like to
emphasize the phases investigated in our paper are not involving these
special points. We also provide some illustrative simulations to show that
the amplitude control of a wave packet can be accomplished by a
quasi-adiabatic process within a short time.

This paper is organized as follows. In Sec.\ \ref{Model and solutions}, we
introduce the non-Hermitian Rice-Mele model and its exact solution. Section %
\ref{Amplitude modulation by imaginary Berry phase} is dedicated to show how
the amplitude modulation is determined by the imaginary Berry phase under
the time-dependent Hamiltonian. In Sec.\ \ref{Wave packet dynamics}, we
investigate wave packet dynamics with its accomplishment of amplitude
control by a quasi-adiabatic process within a relatively short time.
Finally, we give a summary and discussion in Sec.\ \ref{Summary}.

\section{Model and solutions}

\label{Model and solutions}
\begin{figure}[tbp]
\centering
\includegraphics[ bb=1 345 385 729, width=0.35\textwidth, clip]{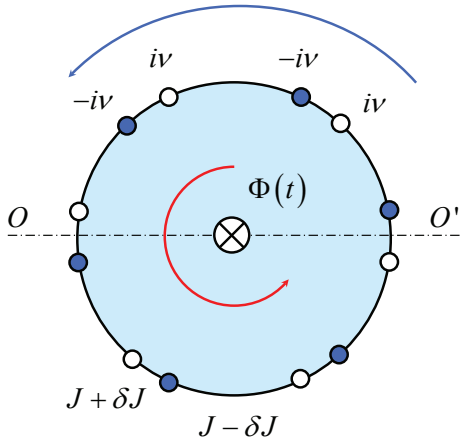}
\par
\caption{(Color online) Schematic illustration of the non-Hermitian
Rice-Mele model driven by a time-dependent external field. It is $\mathcal{PT%
}$-symmetric with respect to the $OO^{\prime }$ axis in the absence of the
field. Constant field $\Phi $\ breaks the $\mathcal{PT}$-symmetry, but keeps
the translational symmetry. A time-varying field $\Phi \left( t\right) $\
induces eddy field in a direction indicated by the red arrow. Together with
the distortion, the imaginary potentials can also break the left-right
chiral symmetry, inducing a direction of the system indicated by the blue
arrow. In the case that two arrows are either the same or the opposite, the
dynamics of a state exhibits different behaviors.}
\label{fig1}
\end{figure}
We consider a non-Hermitian Rice-Mele model \cite{Rice} with a flux, which
can be described by the following Hamiltonian
\begin{eqnarray}
H &=&-J\sum_{j=1}^{2N}\left[ 1+\left( -1\right) ^{j}\delta \right] \left(
e^{i\phi }c_{j}^{\dag }c_{j+1}+\text{\textrm{H.c.}}\right)  \label{H} \\
&&+J\left( \mu +i\nu \right) \sum_{j}\left( -1\right) ^{j}c_{j}^{\dag }c_{j},
\notag
\end{eqnarray}%
on a $2N$-site lattice, where $c_{j}^{\dag }$ is the creation operator of a
boson (or a fermion) at the $j$th site with the periodic boundary condition $%
c_{2N+1}=c_{1}$. Here the hopping amplitude is modulated by the dimerization
factor $\delta $, and $\Phi =2N\phi $ is the magnetic flux threading the
ring. Staggered complex potential induces the non-Hermiticity of the model.

There are three elements in the structure of the model, lattice distortion,
imaginary potentials, and flux. As is shown in Fig. \ref{fig1}, the
imaginary potentials and distortion can break the left-right chiral symmetry
\cite{HWH1}. In addition, a time-varying field $\Phi \left( t\right)$
induces eddy field in another direction. The dynamics of a state should
exhibit different behaviors with different configurations. It turns out that
imaginary potentials can appear in open physical systems \cite%
{Dalibard,Muller,Dum,Plenio,Daley}. In experiments, the effective magnetic
flux threading a ring can be realized by rotating the lattice \cite%
{Tung,Williams}.

In the absence of the imaginary potential and flux, this model has been
adequately studied in various perspectives \cite{Xiao}. In the case of
purely imaginary potential and $\phi =0$, the model has $\mathcal{PT}$
symmetry, and the dynamics has been systematically investigated \cite{HWH1}
in the frameworks of biorthogonal and Dirac inner products. In this paper,
we refer particle probability to Dirac probability. Moreover, we introduce
the magnetic flux to the Rice-Mele model, which has been employed to control
the dynamics of wave packet \cite{Osborne,YS, HWH2}. Although non-zero $\phi
$ breaks the $\mathcal{PT}$ symmetry, we will show that this model can have
full real spectrum. We note that the Hamiltonian is invariant via a
translational transformation, i.e., $\left[ T_{2},H\right] =0$, where $T_{2}$
is the shift operator that defined as
\begin{equation}
T_{2}^{-1}c_{i}^{\dagger }T_{2}=c_{i+2}^{\dagger }\text{.}
\end{equation}%
This allows the invariant subspace spanned by the eigenvector of operator $%
T_{2}$. The single-particle eigenvector of $T_{2}$\ can be expressed as $%
a_{k}^{\dag }\left\vert 0\right\rangle $ and $b_{k}^{\dag }\left\vert
0\right\rangle $, where%
\begin{eqnarray}
a_{k}^{\dag } &=&\frac{1}{\sqrt{N}}\sum_{j}e^{ikj}c_{2j-1}^{\dag }, \\
b_{k}^{\dagger } &=&\frac{1}{\sqrt{N}}\sum_{j}e^{ikj}c_{2j}^{\dag },
\end{eqnarray}%
satisfying%
\begin{equation}
T_{2}^{-1}a_{k}^{\dagger }T_{2}=e^{-ik}a_{k}^{\dagger }\text{, }%
T_{2}^{-1}b_{k}^{\dagger }T_{2}=e^{-ik}b_{k}^{\dagger }\text{.}
\end{equation}%
Here, $a_{k}^{\dag }$ and $b_{k}^{\dagger }$ are two kinds of creation
operators of bosons (or fermions), with $k=2\pi n/N$ ($n\in \lbrack 1,N]$),
representing the particles in odd and even sublattices. Then the original
Hamiltonian $H$ can be expressed as

\begin{equation}
H=\sum_{k}H_{k},
\end{equation}%
where%
\begin{eqnarray}
H_{k} &=&\Lambda \left( k,\phi \right) Ja_{k}^{\dagger }b_{k}+\text{\textrm{%
H.c.}}  \notag \\
&&-\left( \mu +i\nu \right) J\left( a_{k}^{\dagger }a_{k}-b_{k}^{\dagger
}b_{k}\right) ,
\end{eqnarray}%
and%
\begin{equation}
\Lambda \left( k,\phi \right) =-e^{-ik/2}\sum_{\lambda =\pm }\left(
1-\lambda \delta \right) e^{i\lambda \left( k/2+\phi \right) }.
\end{equation}%
It is easy to check that
\begin{equation}
\left[ H_{k},H_{k^{^{\prime }}}\right] =0,
\end{equation}%
which ensures us to arrive at the solution in each invariant subspace.

Considering the single-particle solution, we can introduce the pseudo-spin
operators%
\begin{equation}
s_{k}^{+}=\left( s_{k}^{-}\right) ^{\dag }=a_{k}^{\dagger }b_{k}\text{, }%
s_{k}^{z}=\frac{1}{2}\left( a_{k}^{\dagger }a_{k}-b_{k}^{\dagger
}b_{k}\right) ,
\end{equation}%
which obey
\begin{equation}
\left[ s_{k}^{+},s_{k}^{-}\right] =2s_{k}^{z},\left[ s_{k}^{z},s_{k}^{\pm }%
\right] =\pm s_{k}^{\pm }.
\end{equation}%
Accordingly, $H_{k}$ has the form
\begin{equation}
H_{k}=\vec{B}_{k}\cdot \vec{\sigma}_{k},
\end{equation}%
where $\vec{\sigma}_{k}$ is Pauli matrix. Components of the field $\vec{B}%
_{k}$ in the rectangular coordinates are%
\begin{eqnarray}
B_{k}^{x}/J &=&-\left( 1-\delta \right) \cos \phi -\left( 1+\delta \right)
\cos \left( k+\phi \right) , \\
B_{k}^{y}/J &=&\left( 1-\delta \right) \sin \phi -\left( 1+\delta \right)
\sin \left( k+\phi \right) , \\
B_{k}^{z}/J &=&-\left( \mu +i\nu \right) ,
\end{eqnarray}%
where%
\begin{equation}
\cos \theta _{k}=\frac{B_{k}^{z}}{B_{k}},\tan \varphi _{k}=\frac{B_{k}^{y}}{%
B_{k}^{x}},
\end{equation}%
and the field magnitude is%
\begin{equation*}
B_{k}=[\left( B_{k}^{x}\right) ^{2}+\left( B_{k}^{y}\right) ^{2}+\left(
B_{k}^{z}\right) ^{2}]^{1/2}.
\end{equation*}%
Obviously, $\theta _{k}$\ can be a complex number even in the case with real
$B_{k}$.

The eigenvalues of $H_{k}$ are%
\begin{eqnarray}
\varepsilon _{\pm }^{k} &=&\pm B_{k} \\
&=&\pm 2J[\left( \mu +i\nu \right) ^{2}/4+\delta ^{2}  \notag \\
&&+\left( 1-\delta ^{2}\right) \cos ^{2}\left( k/2+\phi \right) ]^{\frac{1}{2%
}},  \notag
\end{eqnarray}%
which give the spectrum of the whole system when all possible $k$ are taken.
Moreover, the eigenstates of a non-Hermitian Hamiltonian can construct a set
of biorthogonal bases in association with the eigenstates of its Hermitian
conjugate. For the present system, eigenstates $\left\vert \psi
_{+}^{k}\right\rangle $, $\left\vert \psi _{-}^{k}\right\rangle $ of $H_{k}$
and $\left\vert \eta _{+}^{k}\right\rangle $, $\left\vert \eta
_{-}^{k}\right\rangle $ of $H_{k}^{\dagger }$ are the biorthogonal bases of
the single-particle invariant subspace, which are explicitly expressed as,

\begin{equation}
\begin{array}{cc}
\left\vert \psi _{+}^{k}\right\rangle =\left(
\begin{array}{c}
\cos \frac{\theta _{k}}{2} \\
\sin \frac{\theta _{k}}{2}e^{i\varphi _{k}}%
\end{array}%
\right) , & \left\vert \psi _{-}^{k}\right\rangle =\left(
\begin{array}{c}
-\sin \frac{\theta _{k}}{2} \\
\cos \frac{\theta _{k}}{2}e^{i\varphi _{k}}%
\end{array}%
\right) , \\
&  \\
\left\vert \eta _{+}^{k}\right\rangle =\left(
\begin{array}{c}
\cos \frac{\theta _{k}}{2} \\
\sin \frac{\theta _{k}}{2}e^{-i\varphi _{k}}%
\end{array}%
\right) ^{\ast }, & \left\vert \eta _{-}^{k}\right\rangle =\left(
\begin{array}{c}
-\sin \frac{\theta _{k}}{2} \\
\cos \frac{\theta _{k}}{2}e^{-i\varphi _{k}}%
\end{array}%
\right) ^{\ast }.%
\end{array}%
\end{equation}%
It is ready to check that the biorthogonal bases $\left\{ \left\vert \psi
_{\lambda }^{k}\right\rangle ,\left\vert \eta _{\lambda }^{k}\right\rangle
\right\} $ ($\lambda =\pm $) obey the biorthogonal and completeness
conditions
\begin{equation}
\left\langle \eta _{\lambda ^{\prime }}^{k^{\prime }}\right. \left\vert \psi
_{\lambda }^{k}\right\rangle =\delta _{\lambda \lambda ^{\prime }}\delta
_{kk^{\prime }},\text{ }\sum_{\lambda ,k}\left\vert \psi _{\lambda
}^{k}\right\rangle \left\langle \eta _{\lambda }^{k}\right\vert =1.
\end{equation}%
These properties are independent of the reality of the spectrum. In this
paper, we focus on the case with full real spectrum. This happens when (i) $%
\nu =0$, the Hamiltonian goes back to a Hermitian system, (ii) $\mu =0$ and $%
\nu <2\delta $. The phase diagram is true for arbitrary value of constant $%
\phi $, which is the basis for the subsequent investigation of the dynamics
with time-dependent flux.

\section{Amplitude modulation by imaginary Berry phase}

\label{Amplitude modulation by imaginary Berry phase}

The aim of this paper is to investigate the effect of the time-varying flux
on the dynamics of the non-Hermitian system. We start with an adiabatic
evolution, in which an initial eigenstate evolves into the instantaneous
eigenstate of the time-dependent Hamiltonian.

From Eq. (\ref{H}), we know that $H$ is a periodic function of $\phi $ with $%
H\left( \phi \right) $ $=H\left( \phi +2\pi \right) $ and $H_{k}\left( \phi
\right) $ $=H_{k}\left( \phi +2\pi \right) $. Considering the time-dependent
flux $\phi \left( t\right) $, any eigenstate $\left\vert \psi _{\lambda
}^{k}\left( 0\right) \right\rangle $\ will return back to $\left\vert \psi
_{\lambda }^{k}\left( 0\right) \right\rangle $\ if $\phi \left( t\right) $\
varies adiabatically from $0$ to $2\pi $, and the evolved state is the
instantaneous eigenstate $\left\vert \psi _{\lambda }^{k}\left( \phi \right)
\right\rangle $. More explicitly, we regard the flux as a linear function of
time, that is, $\phi =\beta t$. And the adiabatic evolution of the initial
eigenstate $\left\vert \psi _{\lambda }^{k}\left( 0\right) \right\rangle $
under the time-dependent Hamiltonian $H\left( \phi \left( t\right) \right) $
can be expressed as
\begin{eqnarray}
\left\vert \Psi _{\lambda }^{k}\left( \phi \right) \right\rangle &=&\mathcal{%
T}\exp \left[ -i\int_{0}^{t}H\left( t\right) \mathrm{d}t\right] \left\vert
\psi _{\lambda }^{k}\left( 0\right) \right\rangle \\
&=&e^{i\left( \alpha _{k}^{\lambda }+\gamma _{k}^{\lambda }\right)
}\left\vert \psi _{\lambda }^{k}\left( \phi \right) \right\rangle .  \notag
\end{eqnarray}%
Here the dynamics phase defined by $\alpha _{k}^{\lambda }\left( \phi
\right) $ and adiabatic phase $\gamma _{k}^{\lambda }\left( \phi \right) $
have the form
\begin{eqnarray}
\alpha _{k}^{\lambda }\left( \phi \right) &=&-\frac{1}{\beta }%
\int\nolimits_{0}^{\phi }\varepsilon _{\lambda }^{k}\left( \phi \right)
\mathrm{d}\phi ,  \label{Alpha_k} \\
\gamma _{k}^{\lambda }\left( \phi \right) &=&i\int\nolimits_{0}^{\phi
}\left\langle \eta _{\lambda }^{k}\left( \phi \right) \right\vert \partial
_{\phi }\left\vert \psi _{\lambda }^{k}\left( \phi \right) \right\rangle
\mathrm{d}\phi  \label{Gamma_k} \\
&=&-\int\nolimits_{0}^{\phi }\frac{2\delta J^{2}\mathrm{d}\phi }{\varepsilon
_{\lambda }^{k}\left[ \varepsilon _{\lambda }^{k}-J\left( \mu +i\nu \right) %
\right] }.  \notag
\end{eqnarray}%
\begin{figure*}[tbp]
\centering
\includegraphics[ bb=50 117 581 520, width=0.42\textwidth, clip]{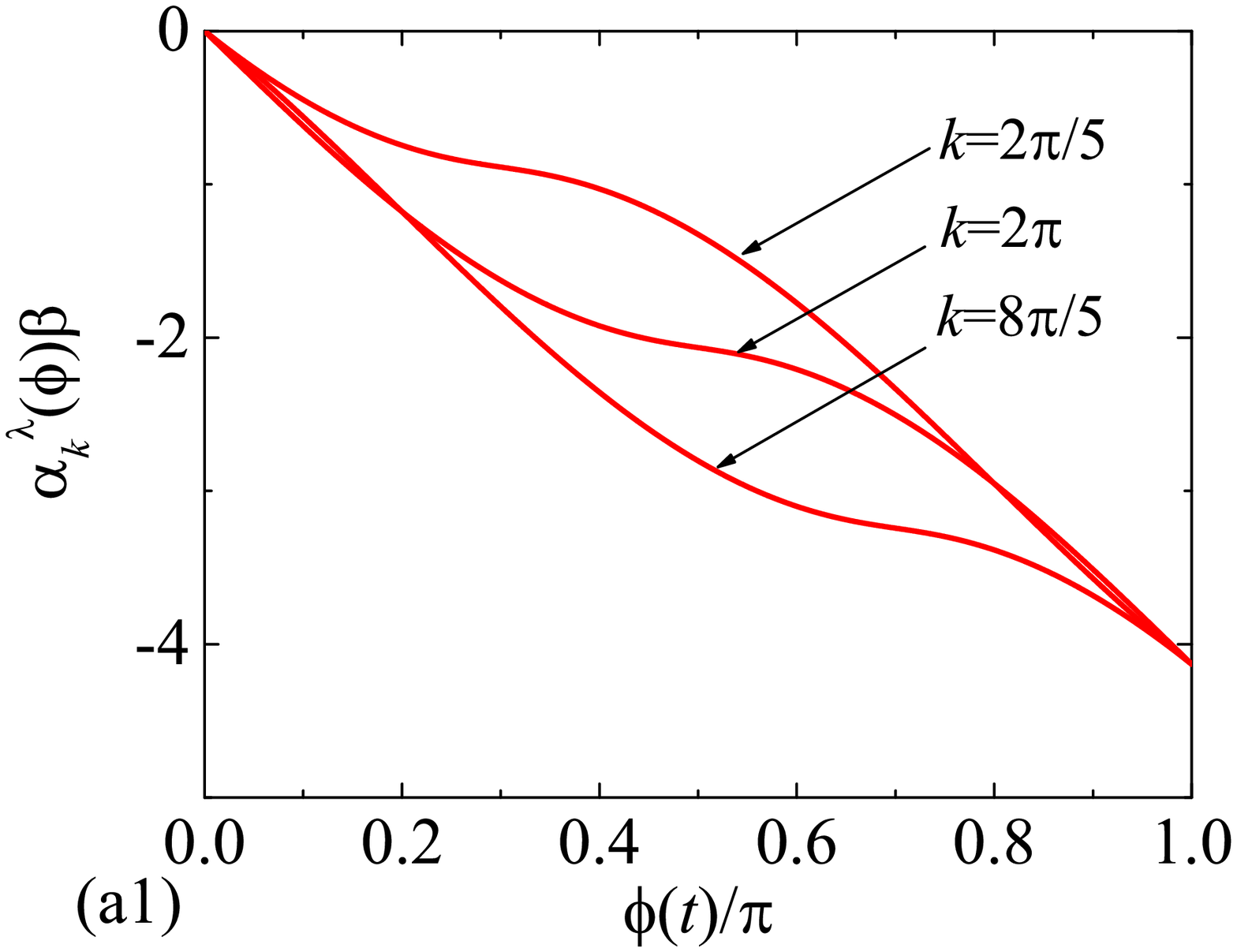} %
\includegraphics[ bb=50 117 581 520, width=0.42\textwidth, clip]{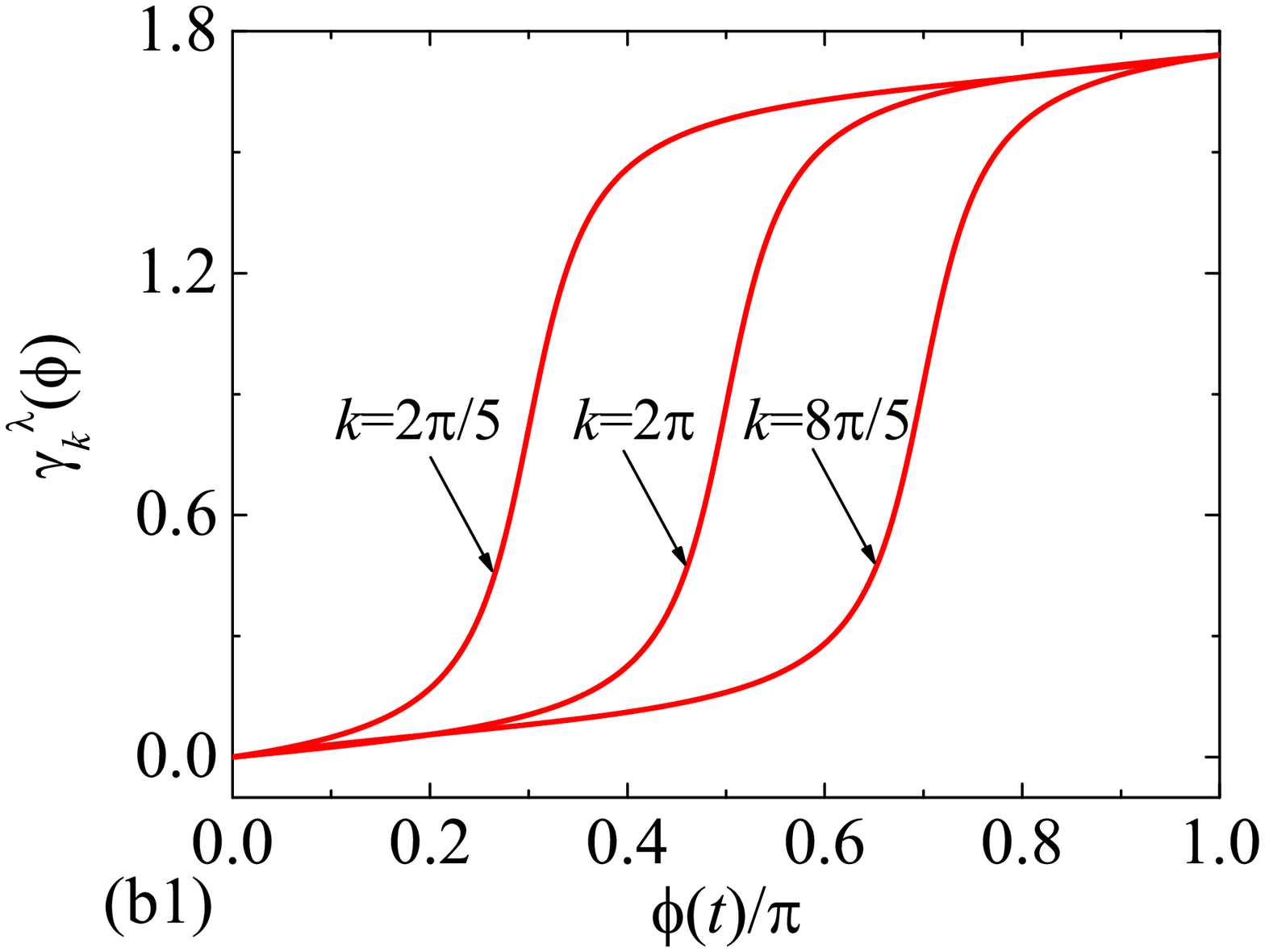} %
\includegraphics[ bb=50 117 581 520, width=0.42\textwidth, clip]{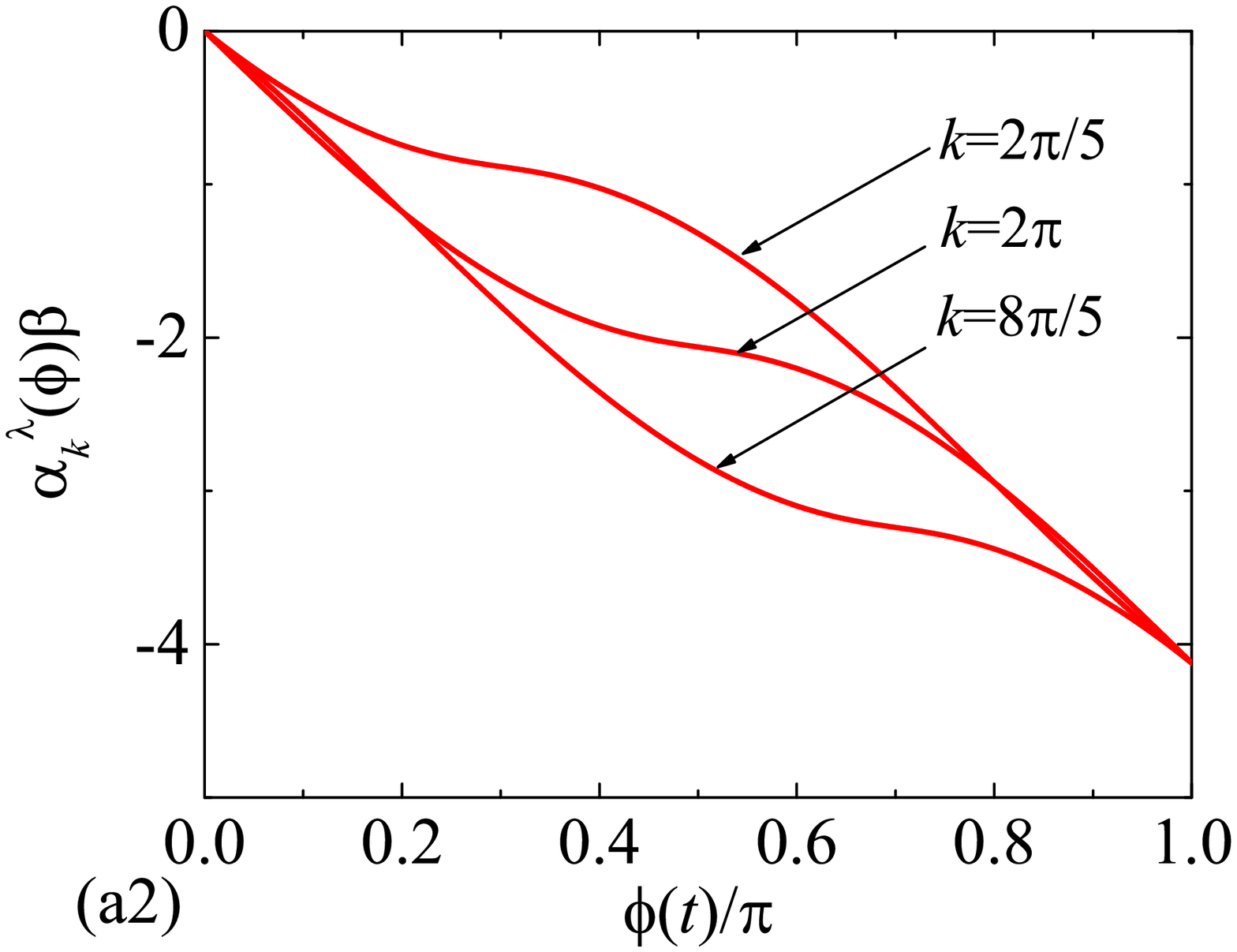} %
\includegraphics[ bb=50 117 581 520, width=0.42\textwidth, clip]{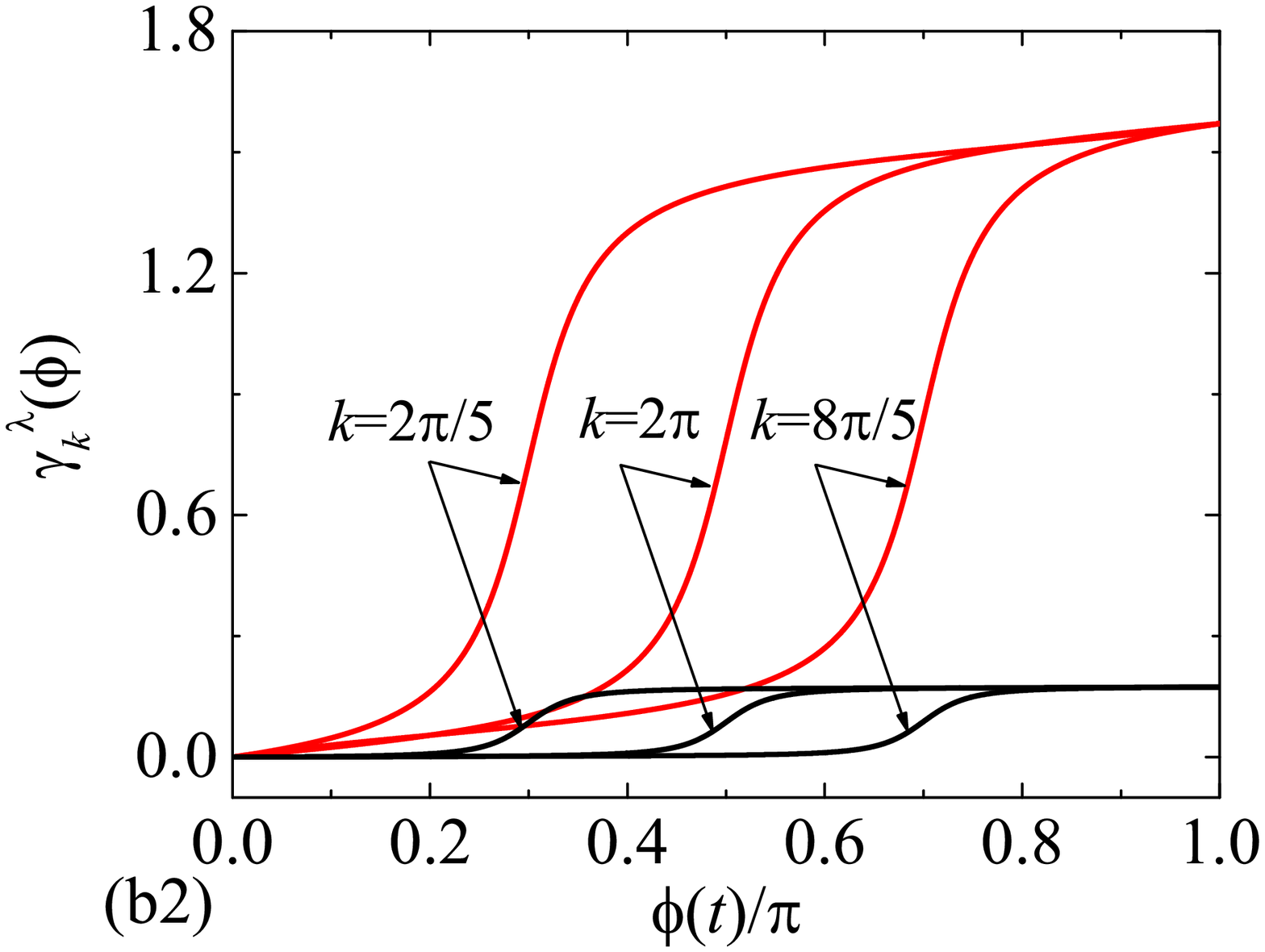}
\par
\caption{(Color online) Plots of the reduced dynamic phase $\protect\alpha %
_{k}^{\protect\lambda }\protect\beta $ and the geometric phase $\protect%
\gamma _{k}^{\protect\lambda }$ as functions of $\protect\phi $, expressed
in Eq. (\protect\ref{Alpha_k}) and (\protect\ref{Gamma_k}), for different
values of $k$ in a typical (a) Hermitian system with $\protect\delta =-0.15$%
, $\protect\mu =0.05$, and $\protect\nu =0$; (b) non-Hermitian system with $%
\protect\delta =-0.15$, $\protect\mu =0$, and $\protect\nu =0.05$. Both
systems have full real spectrum. The real (imaginary) parts of the phases
are plotted in red (black). (a) It is shown that both phases are real for
the Hermitian system, while the adiabatic phase is complex for the
non-Hermitian system. The dynamic phases and the real part of adiabatic
phases in both cases have a slight difference. The shapes of all the phases
are in agreement with the approximate expressions in Eqs. (\protect\ref%
{H_gamma}), (\protect\ref{H_alfa}), (\protect\ref{NH_gamma}), and (\protect
\ref{NH_alfa}). Here the reduced dynamic phase $\protect\alpha _{k}^{\protect%
\lambda }\protect\beta $ and the geometric phase $\protect\gamma _{k}^{%
\protect\lambda }$ are expressed in units of radian.}
\label{fig2}
\end{figure*}
In this paper, we focus on the system with full real spectrum. Then the
dynamics phase depends only on the instantaneous dispersion relation and is
always real. We have an interest in the adiabatic phase since it may have an
extra contribution to an evolved state. In the following, we will present
several\ features of $\gamma _{k}^{\lambda }\left( \phi \right) $, based on
Eq. (\ref{Gamma_k}).

It is obvious that we have $\gamma _{k}^{\lambda }\left( \phi \right) =0$,
if the dimerization factor $\delta =0$, which is a necessary condition for
nonzero adiabatic phase. In contrast, we have nonzero $\gamma _{k}^{\lambda
}\left( \phi \right) $\ in the case of zero staggered potential, which
indicates the significance of\ the Peierls distortion to the adiabatic
phase.\ It is easy to check that

\begin{equation}
\frac{\partial ^{2}}{\partial ^{2}\phi }\gamma _{k}^{\lambda }\left( \phi
\right) \propto \sin \left( k+2\phi \right)
\end{equation}%
which leads to $\partial _{\phi }^{2}\gamma _{k}^{\lambda }\left( \phi
\right) =0$ at $\phi _{c}=\pi /2-k/2$. This fact indicates that both the
real and imaginary parts of $\gamma _{k}^{\lambda }\left( \phi \right) $\
experience a maximal (or minimal) change at this point. Furthermore,
including Hermitian and non-Hermitian systems, we note that%
\begin{equation}
\frac{\partial \gamma _{k}^{\lambda }\left( n\pi \right) }{\partial k}=\frac{%
\partial \alpha _{k}^{\lambda }\left( n\pi \right) }{\partial k}=0.
\end{equation}
In view of the fact that for an arbitrary function $g\left( \cos \left(
k+2\phi \right) \right)$, one can always obtain
\begin{eqnarray}
&&\frac{\partial }{\partial k}\int\nolimits_{0}^{n\pi }g\left( \cos \left(
k+2\phi \right) \right) \mathrm{d}\phi  \notag \\
&=&\frac{1}{2}g\left( \cos \left( k+2\phi \right) \right) _{0}^{n\pi }=0.
\end{eqnarray}%
This shows that the dynamic and Berry phases are $k$-independent after $\phi
$\ varying $n\pi $, which ensures that any arbitrary initial state involved
in the upper or lower band solely can revive back exactly after $\phi $\
varying $2\pi $.

For a Hermitian system, the adiabatic phase is always real, which ensures
the probability preserving evolution. While, the probability of an evolved
state changes due to the imaginary part of the adiabatic phase in a
non-Hermitian system. The gain or loss of probability depends on the sign of
the imaginary phase. There are several rigorous results for the phase.

For given parameters $\{\delta ,\mu ,\nu \}$, the adiabatic phase of an
eigenstate in $\lambda $ band obeys the identity
\begin{equation}
\gamma _{k}^{\lambda }\left( n\pi \right) =-\gamma _{k}^{\lambda }\left(
-n\pi \right) ,  \label{Gama-Gama}
\end{equation}%
which is obtained from Eq. (\ref{Gamma_k}), owing to the fact%
\begin{equation}
\varepsilon _{\lambda }^{k}\left( \phi -n\pi \right) =\varepsilon _{\lambda
}^{k}\left( \phi \right) .  \label{e-e}
\end{equation}%
This means that the direction of $\phi $\ can determine the sign of $\text{Im%
}\gamma _{k}^{\lambda }\left( n\pi \right) $, controlling the probability of
an evolved state.

In addition, the sign of $\text{Re}\gamma _{k}^{\lambda }\left( \phi \right)
$ and $\text{Im}\gamma _{k}^{\lambda }\left( \phi \right) $ could also
depend on the sign of $\delta $ and $\nu $ via the Eqs. (\ref{re}) and (\ref%
{im}), that is,
\begin{equation}
\text{Re}\gamma _{k}^{\lambda }\left( \phi \right) =\text{\textrm{sgn}}%
\left( \delta \right) \int\nolimits_{0}^{\phi }\frac{-2\left\vert \delta
\right\vert J^{2}\mathrm{d}\phi }{\left[ \left( B_{k}\right) ^{2}+J^{2}\nu
^{2}\right] },  \label{re}
\end{equation}

\begin{equation}
\text{Im}\gamma _{k}^{\lambda }\left( \phi \right) =\text{\textrm{sgn}}%
\left( \delta \nu \lambda \right) \int\nolimits_{0}^{\phi }\frac{%
-2J^{3}\left\vert \delta \nu \right\vert \mathrm{d}\phi }{B_{k}\left[ \left(
B_{k}\right) ^{2}+J^{2}\nu ^{2}\right] },  \label{im}
\end{equation}%
which yield
\begin{eqnarray}
\left[ \gamma _{k}^{\lambda }\left( \phi \right) \right] _{\delta } &=&-%
\left[ \gamma _{k}^{\lambda }\left( \phi \right) \right] _{-\delta }, \\
\left[ \gamma _{k}^{\lambda }\left( \phi \right) \right] _{\nu } &=&\left[
\gamma _{k}^{\lambda }\left( \phi \right) \right] _{-\nu }^{\ast }.
\end{eqnarray}%
Together with the Eq. (\ref{Gama-Gama}), we show that the sign of $\text{%
\textrm{Im}}\left( \gamma _{k}^{\lambda }\right) $ is determined by the
following expression that
\begin{equation}
\text{sgn}\left[ \text{\textrm{Im}}\gamma _{k}^{\lambda }\left( \lambda
^{\prime }n\pi \right) \right] =-\text{sgn}\left( \nu \delta \lambda \lambda
^{\prime }\right) \quad \left( \lambda ^{\prime }=\pm \right) ,
\label{total sign}
\end{equation}%
which directly results in the amplification or attenuation of an evolved
state.

Besides the exact results, it is useful to arrive at the whole profile of
phases as a function of a group of parameters $\{\phi ,k,\delta ,\mu ,\nu
,\lambda \}$. The approximate expressions of the phases for Hermitian and
non-Hermitian systems could be obtained from straightforward derivations as
following, respectively.

For a Hermitian system $\left( \nu =0\right) $, we have%
\begin{eqnarray}
&&\gamma _{k}^{\lambda }\left( \phi \right) \approx \frac{\text{sgn}\left(
\delta /2\right) }{\sqrt{1-\delta ^{2}}}\{\tan ^{-1}\Theta _{k}\left(
0\right) -\tan ^{-1}\Theta _{k}\left( \phi \right)  \label{H_gamma} \\
&&-\text{sgn}\left( \mu \lambda \right) \left[ \tan ^{-1}\left( \left\vert
\mu \right\vert \Gamma _{k}\left( \phi \right) \right) -\tan ^{-1}\left(
\left\vert \mu \right\vert \Gamma _{k}\left( 0\right) \right) \right] \},
\notag
\end{eqnarray}%
and
\begin{eqnarray}
&&\alpha _{k}^{\lambda }\left( \phi \right) \approx \frac{\lambda }{4\beta }%
\{B_{k}\left( 0\right) \left( k-\pi \right) -B_{k}\left( \phi \right) \left(
k+2\phi -\pi \right)  \notag \\
&&+\frac{J\left( 4\delta ^{2}+\mu ^{2}\right) }{\sqrt{1-\delta ^{2}}}\ln
\frac{B_{k}\left( \phi \right) -2\left\vert \delta \right\vert J\Theta
_{k}\left( \phi \right) }{B_{k}\left( 0\right) -2\left\vert \delta
\right\vert J\Theta _{k}\left( 0\right) }\},  \label{H_alfa}
\end{eqnarray}%
where%
\begin{eqnarray}
\Theta _{k}\left( \phi \right) &=&\sqrt{\delta ^{-2}-1}\left( k+2\phi -\pi
\right) /2, \\
\Gamma _{k}\left( \phi \right) &=&J\Theta _{k}\left( \phi \right)
/B_{k}\left( \phi \right) ,
\end{eqnarray}%
are even functions of $\delta ,\mu ,$ and $\lambda $. It is shown that $%
\gamma _{k}^{\lambda }\left( \phi \right) $\ is a real number, preserving
the probability.

While, for the non-Hermitian system $\left( \mu =0,\nu <2\delta \right) $,
we have
\begin{eqnarray}
&&\gamma _{k}^{\lambda }\left( \phi \right) \approx \frac{\text{sgn}\left(
\delta /2\right) }{\sqrt{1-\delta ^{2}}}\{\tan ^{-1}\Theta _{k}\left(
0\right) -\tan ^{-1}\Theta _{k}\left( \phi \right)  \label{NH_gamma} \\
&&-i\text{sgn}\left( \nu \lambda \right) [\tanh ^{-1}\left( \left\vert \nu
\right\vert \Gamma _{k}\left( \phi \right) \right) -\tanh ^{-1}\left(
\left\vert \nu \right\vert \Gamma _{k}\left( 0\right) \right) ]\},  \notag
\end{eqnarray}%
and%
\begin{eqnarray}
&&\alpha _{k}^{\lambda }\left( \phi \right) \approx \frac{\lambda }{4\beta }%
\{B_{k}\left( 0\right) \left( k-\pi \right) -B_{k}\left( \phi \right) \left(
k+2\phi -\pi \right)  \notag \\
&&+\frac{J\left( 4\delta ^{2}-\nu ^{2}\right) }{\sqrt{1-\delta ^{2}}}\ln
\frac{B_{k}\left( \phi \right) -2\left\vert \delta \right\vert J\Theta
_{k}\left( \phi \right) }{B_{k}\left( 0\right) -2\left\vert \delta
\right\vert J\Theta _{k}\left( 0\right) }\},  \label{NH_alfa}
\end{eqnarray}%
where $\Theta _{k}\left( \phi \right) $\ and\ $\Gamma _{k}\left( \phi
\right) $ have the same form as above, but even functions of $\delta ,\nu ,$
and $\lambda $. The remarkable feature of $\gamma _{k}^{\lambda }$ is that
it\ is a complex number.\ The approximate expression shows that both real
and imaginary parts of $\gamma _{k}^{\lambda }\left( \phi \right) $\ are
flat functions of $\phi $ except for the region around the point $\phi _{c}$%
, in which they experience a drastic change. The key feature of an imaginary
phase is its sign, which affects the amplitude of the evolved eigenstate
directly, determining the gain or loss of the probability.

\begin{figure}[tbp]
\centering
\includegraphics[ bb=25 117 581 520, width=0.42\textwidth, clip]{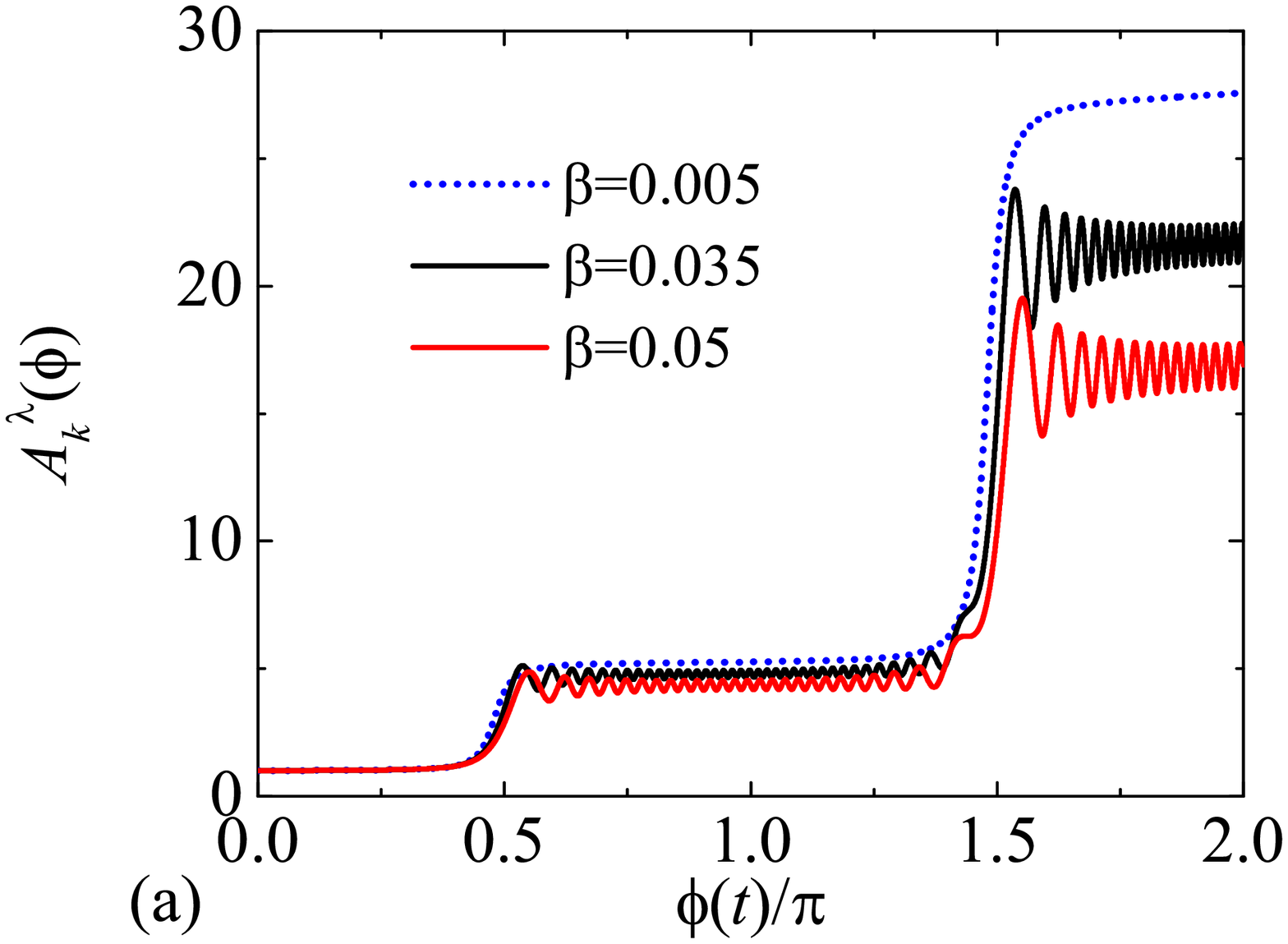} %
\includegraphics[ bb=25 117 581 520, width=0.42\textwidth, clip]{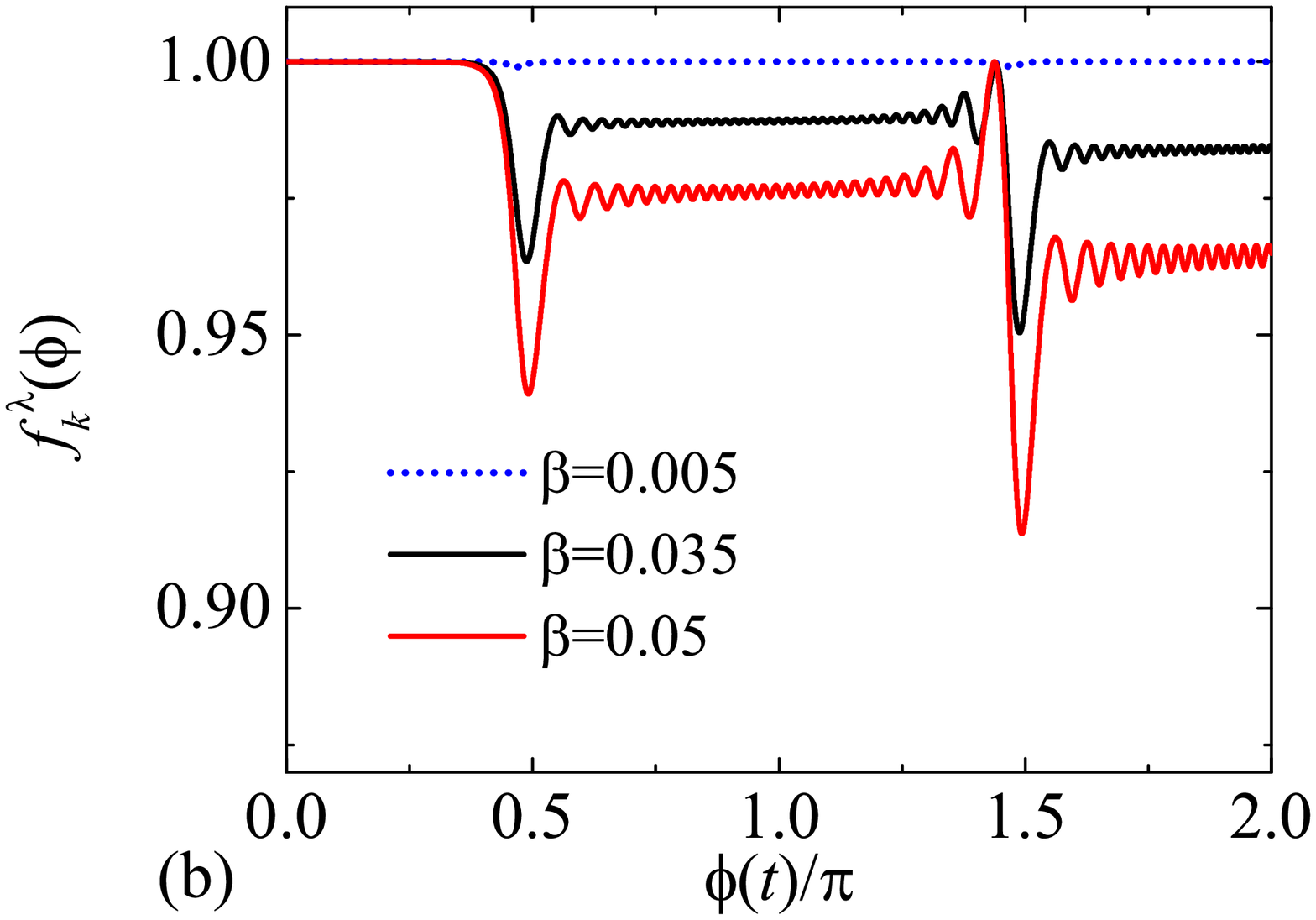}
\par
\caption{(Color online) Numerical simulations for the time evolutions with
different $\protect\beta $. Amplification factor $A_{k}^{\protect\lambda %
}\left( \protect\phi \right) $ and fidelity $f_{k}^{\protect\lambda }\left(
\protect\phi \right) $ for the initial state with $k=\protect\pi /25$ and $%
\protect\lambda =-$ are plotted for the system with $\protect\delta =0.15$, $%
\protect\mu =0$, $\protect\nu =-0.2$, and $N=50$. We can see the evolution
results are close to that, $A_{\protect\pi /25}^{-}\left( \protect\pi %
\right) =27.58,f_{\protect\pi /25}^{-}\left( \protect\pi \right) =1,$
obtained in the adiabatic limit. This indicates that the amplitude control
could be realized with a high fidelity via a quasi-adiabatic process. }
\label{fig3}
\end{figure}

To verify and demonstrate the above analysis, numerical simulations are
performed to investigate the dynamics behavior of a quasi-adiabatic process.
We compute the time evolution of an eigenstate by using a uniform mesh in
the time discretization for the time-dependent Hamiltonian $H(t)$. The
amplification factor (gain) is defined as%
\begin{equation}
A_{k}^{\lambda}\left( \phi \right) =\left\vert \left\vert \psi _{\lambda
}^{k}\left( 0\right) \right\rangle \right\vert ^{-1}\left\vert \mathcal{T}%
\exp \left[ -i\int_{0}^{t}H\left( t\right) \mathrm{d}t\right] \left\vert
\psi _{\lambda }^{k}\left( 0\right) \right\rangle \right\vert ,
\end{equation}%
which is the ratio of the output magnitude to the input magnitude of an
eigenstate. We use the fidelity $f_{k}^{\lambda}\left( \phi \right) $, which
is defined as%
\begin{equation}
f_{k}^{\lambda}\left( \phi \right) =\left\vert \left\langle \Psi _{\lambda
}^{k}\left( \phi \right) \right\vert \mathcal{T}\exp \left[
-i\int_{0}^{t}H\left( t\right) \mathrm{d}t\right] \left\vert \psi _{\lambda
}^{k}\left( 0\right) \right\rangle \right\vert ,
\end{equation}%
to describe the derivation between adiabatic and quasi-adiabatic processes.
For an adiabatic process $\left( \beta \rightarrow 0\right) $, we have $%
A_{k}^{\lambda}\left( \phi \right) =\exp \left[ -\text{\textrm{Im}}\left(
\gamma _{k}^{\lambda }\left( \phi \right) \right) \right] $ and $%
f_{k}^{\lambda}\left( \phi \right) =1$. We can employ $A_{k}^{\lambda}\left(
\pi \right) $ to describe the amplification factor for an arbitrary quantum
state. The computation is performed by using a uniform mesh in the time
discretization for the time-dependent Hamiltonian $H(t)$. As an example, in
Fig. \ref{fig3}, we show the evolution of $A_{k}^{\lambda}\left( \phi
\right) $ and $f_{k}^{\lambda}\left( \phi \right) $\ for different values of
$\beta $. The plot in (a) and (b) shows the quasi-adiabatic process can be
close to the adiabatic one.

\section{Wave packet dynamics}

\label{Wave packet dynamics}
\begin{figure}[tbp]
\centering
\includegraphics[ bb=4 3 602 585, width=0.45\textwidth, clip]{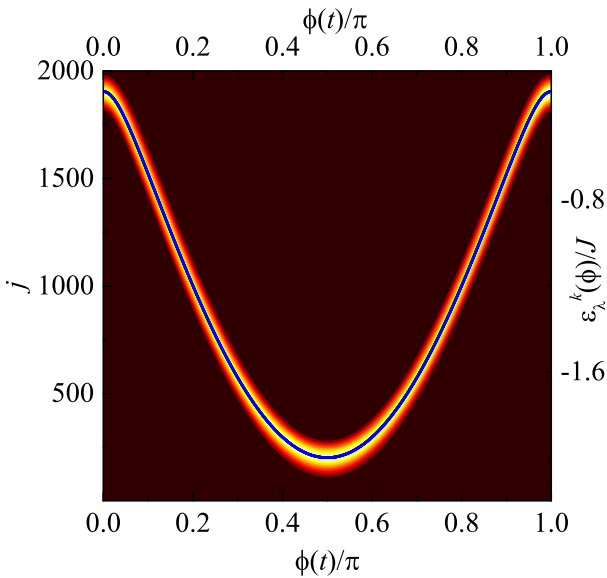}
\par
\caption{(Color online) The comparison between the dispersion relation (blue
dot) and the trajectory of the center of the wave packet (color contour
map). The simulation is computed as follows: the wave packet with $k_{0}=%
\protect\pi /2$, $\protect\alpha =0.05$, and $N_{\mathrm{A}}=1900$, in the
lower band of the system with $\protect\delta =-0.15$, $\protect\mu =0$, $%
\protect\nu =0.05$, and $N=1000$, subjected in the external field with $%
\protect\beta =0.001$. This shows that the two are in close proximity owing
to the fact that the contribution of the geometric phase to the wave packet
position becomes negligible when compared with the dynamic phase. Only the
imaginary part of the geometric phase takes an important role in the
dynamics of the wave packet.}
\label{fig4}
\end{figure}
\begin{figure*}[tbp]
\centering
\includegraphics[ bb=0 15 594 566, width=0.32\textwidth, clip]{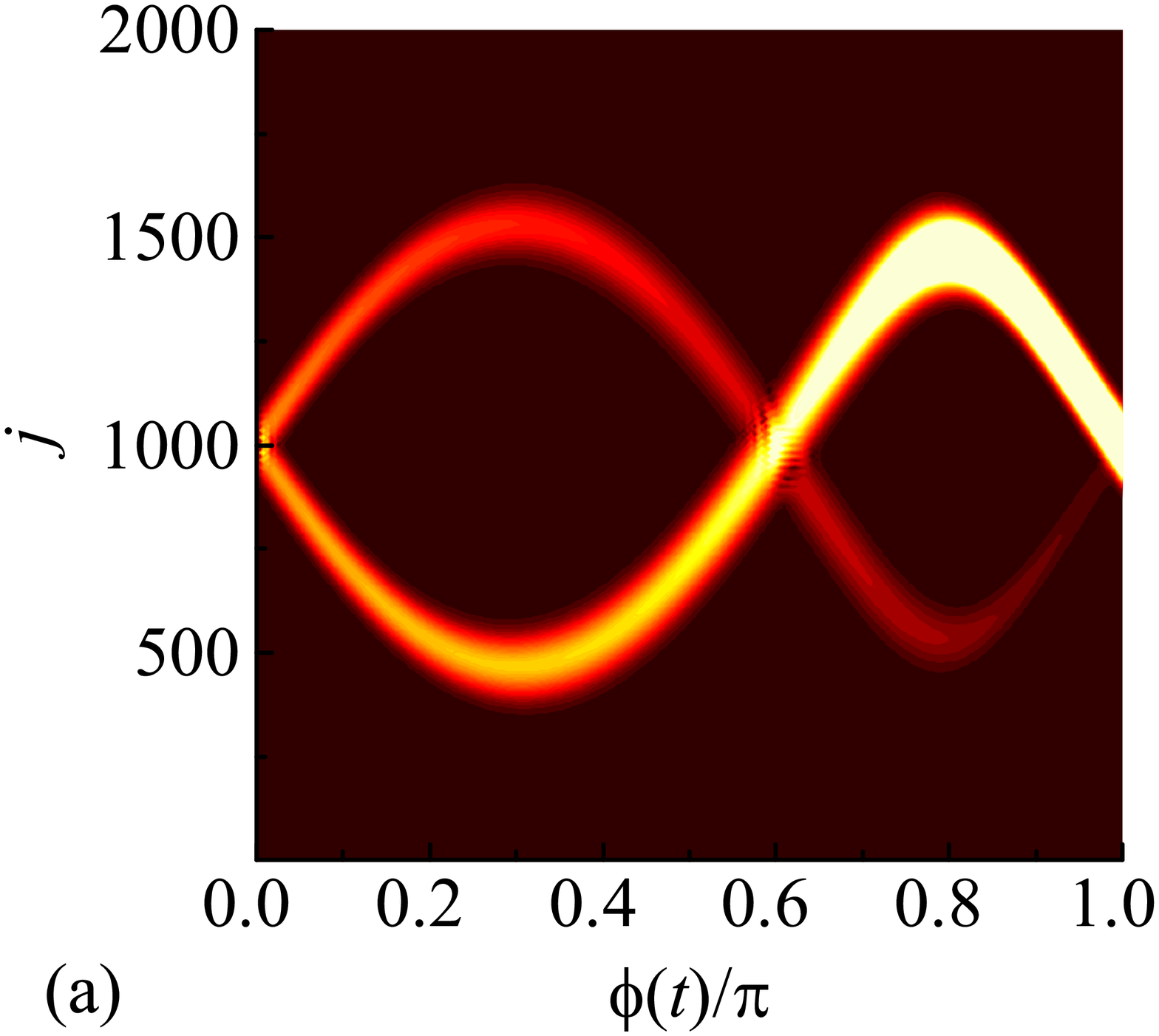} %
\includegraphics[ bb=0 15 594 566, width=0.32\textwidth, clip]{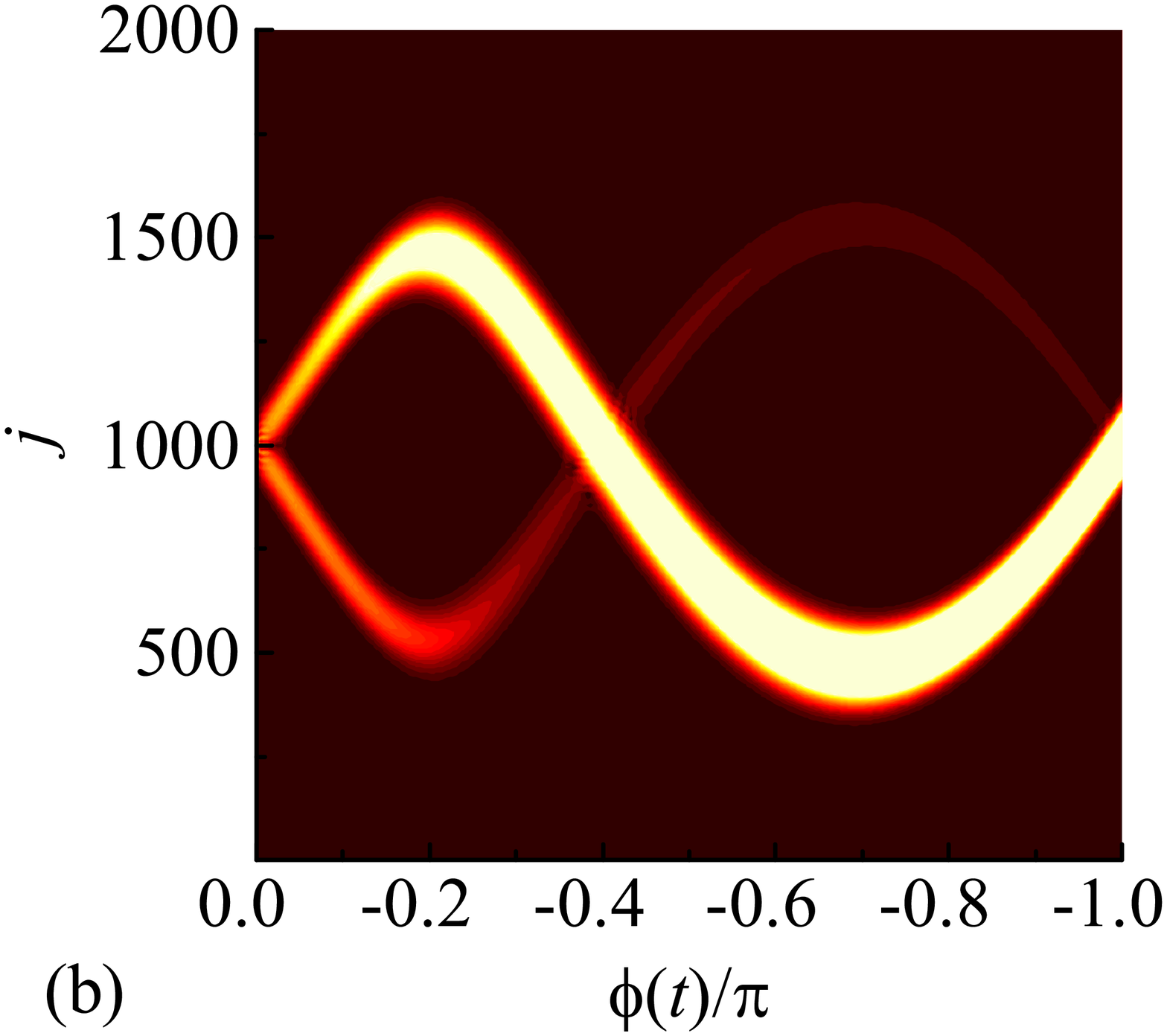} %
\includegraphics[ bb=0 15 594 566, width=0.32\textwidth, clip]{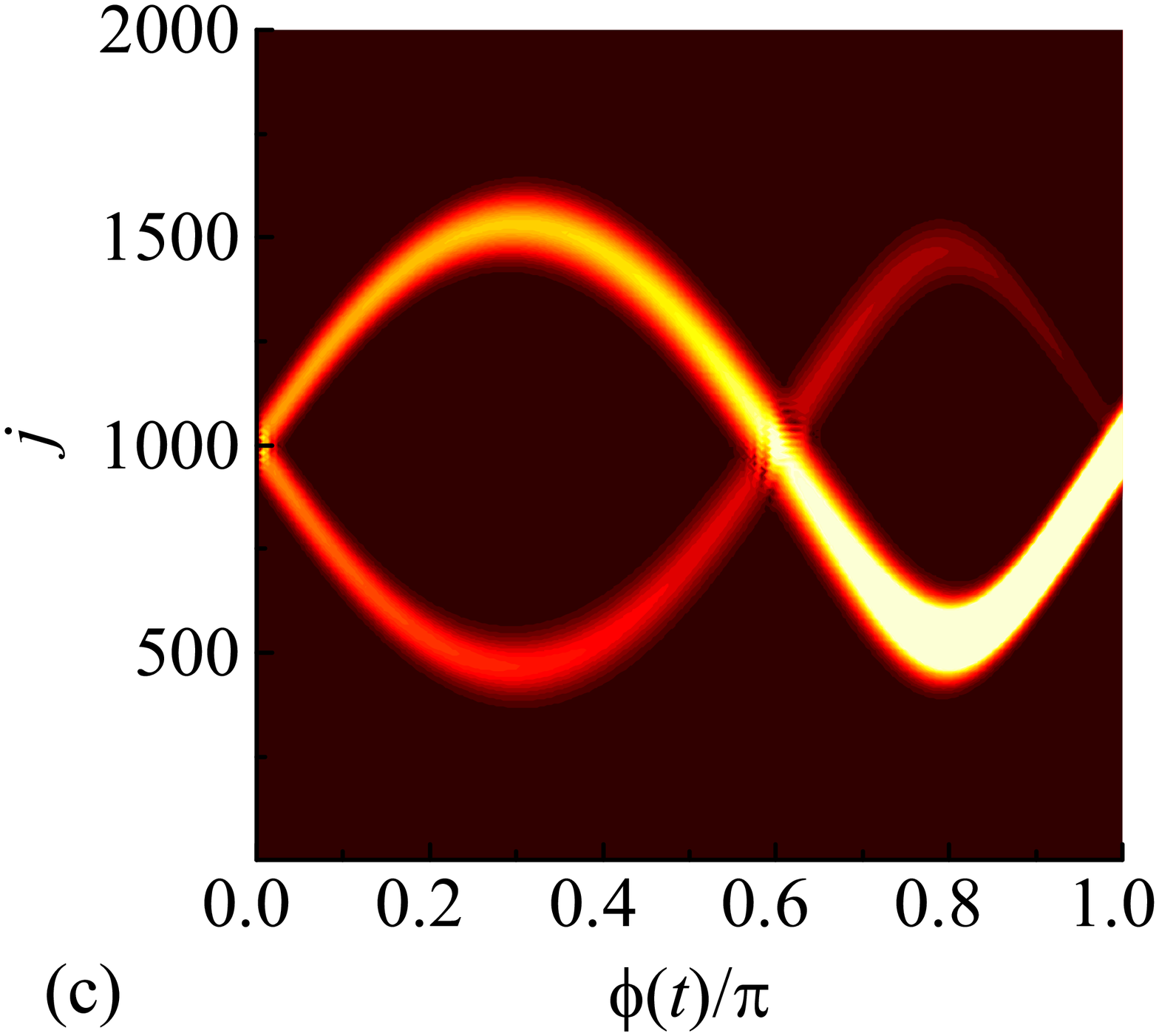}
\par
\caption{(Color online) The profiles of the time evolution of a wave packet
in several typical cases. The initial wave packet is in the form of Eq. (%
\protect\ref{gaussian}) with $k_{0}=1.4\protect\pi $, $\protect\alpha =0.05$%
, and $N_{\mathrm{A}}=1000$, and the external field varies with $\protect%
\beta =0.001$. (a) In the case of $\protect\delta =-0.7$, $\protect\mu =0$,
and $\protect\nu =1.3$, the plot is shown that the probability of the
sub-wave packet in one band increases but the one in another band decreases.
In the cases of (b) and (c), we take the same parameters with (a) but
opposite $\protect\phi $\ and $\protect\delta $, respectively. Comparing
with the profile in (a), we can see the completely opposite behaviors in (b)
and (c). This is in agreement with our analytical prediction.}
\label{fig5}
\end{figure*}
\begin{figure*}[tbp]
\centering
\includegraphics[ bb=72 117 581 520, width=0.325\textwidth, clip]{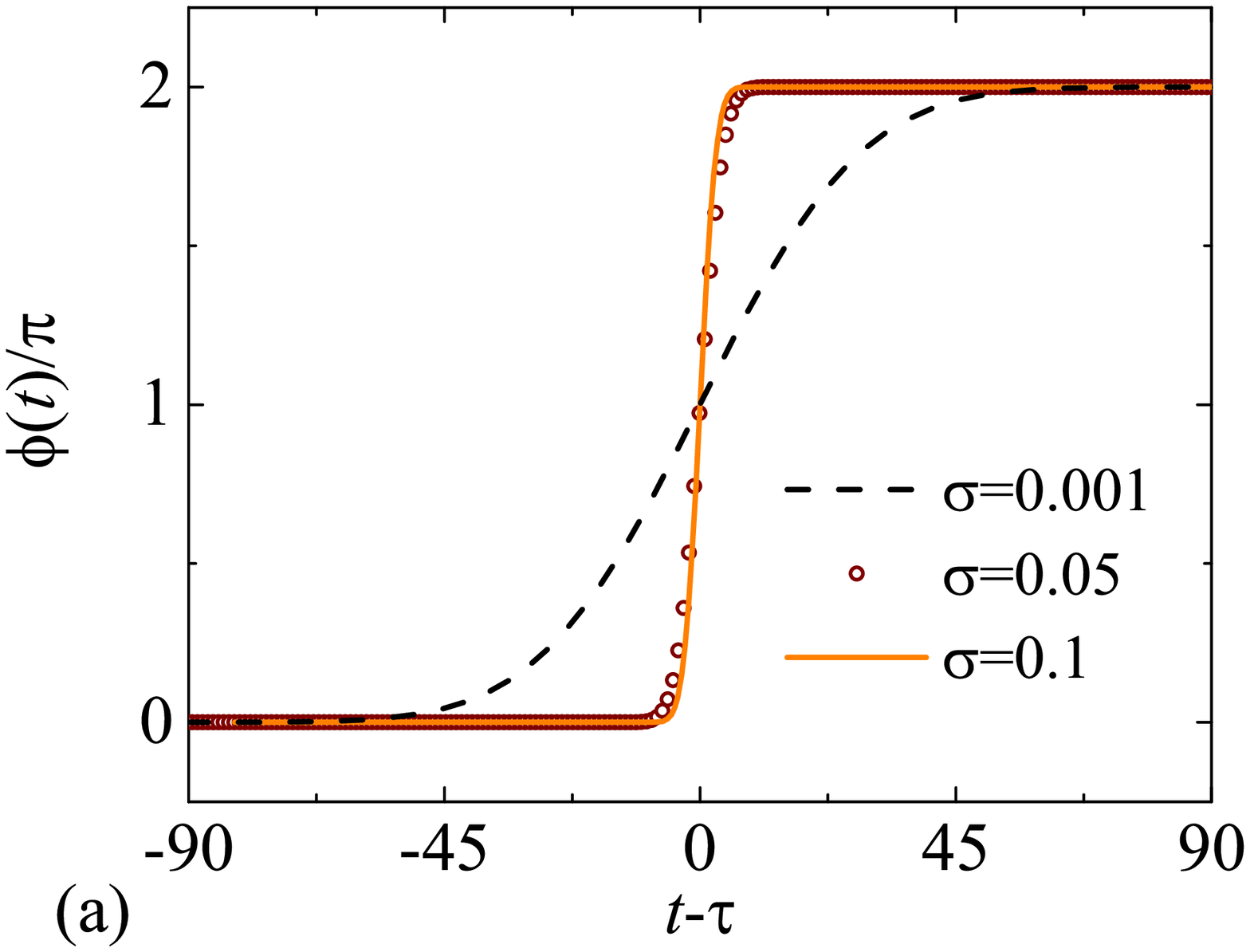} %
\includegraphics[ bb=72 117 581 520, width=0.325\textwidth, clip]{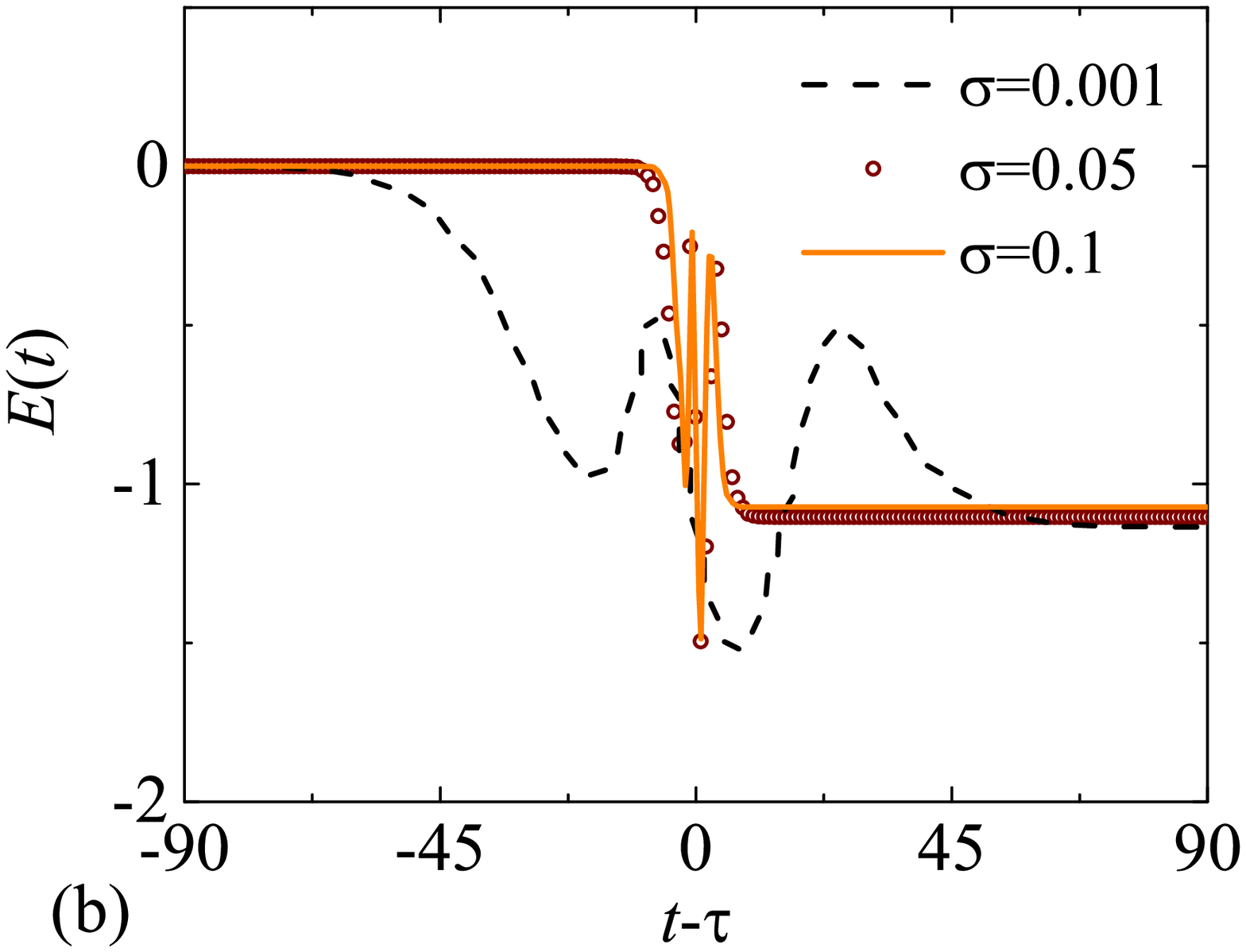} %
\includegraphics[ bb=72 117 581 520, width=0.325\textwidth, clip]{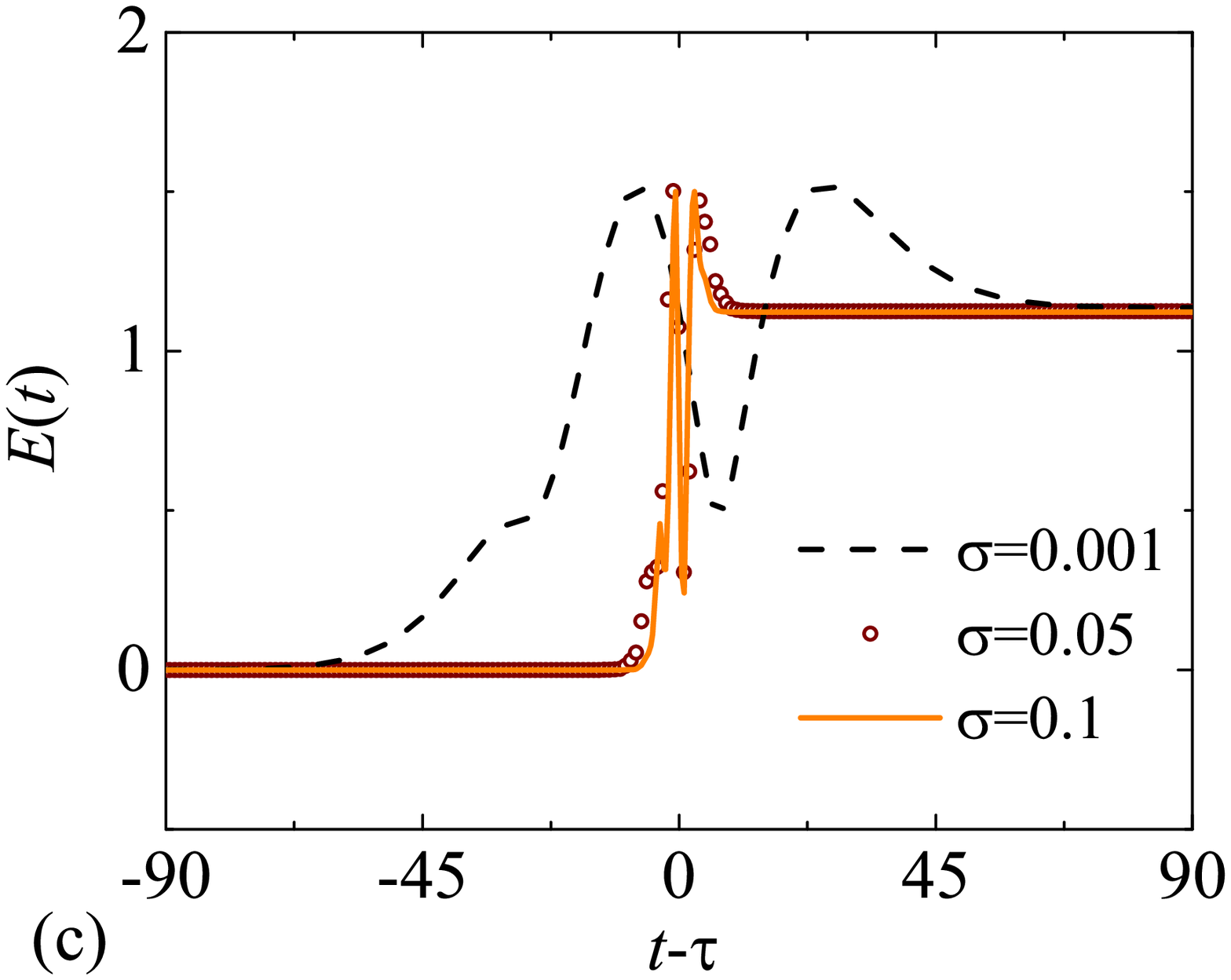}
\par
\caption{(Color online) Profiles of the reduced energy defined in Eq. (%
\protect\ref{ET}) of an evolved wave packet in Eq. (\protect\ref{gaussian})
is controlled by a Gaussian-type time-varying field in Eq. (\protect\ref%
{phi_t}). The parameters of the wave packet and system are $k_{0}=1.5\protect%
\pi $, $\protect\delta =-0.7$, $\protect\mu =0$, and $\protect\nu =1.3$. (a)
The shapes of the field as a function of time in Eq. (\protect\ref{phi_t})
with different $\protect\sigma $. (b) Plots of $E(t)$ of the field in (a).
(c) Same as (b) but with the opposite flux. The simulation result is in
agreement with the result $E(\infty )=-1.14$ for (b) and 1.14 for (c),
obtained in the adiabatical limit. This indicates that the amplitude of the
wave packet can be well controlled via a quasi-adiabatic process in a short
time. Here the time $t-\protect\tau$ and the reduced energy $E\left( t\right)$
are expressed in units of $1/J$ and $J$, respectively. }
\label{fig6}
\end{figure*}
Before starting the investigation of the wave packet dynamics in the present
system, we would like to give a brief review on the dynamics in a uniform
ring. In the case of $\delta =\mu =\nu =0$, it has been shown that the
dynamics of a wave packet is the same as that driven by a linear field with
strength $\beta $, according to the quantum Faraday's law \cite{HWH2}.
Furthermore, it is turned out that the center path of a wave packet driven
by a linear field accords with the dispersion of the Hamiltonian in the
absence of the field within the adiabatic regime \cite{LS}.

\begin{equation}
x_{c}\left( \phi \right) =x_{c}\left( 0\right) +\frac{1}{\beta }\left\{
\varepsilon ^{k_{c}}\left( \phi \right) -\varepsilon ^{k_{c}}\left( 0\right)
\right\} ,
\end{equation}%
where $\varepsilon ^{k}\left( \phi \right) $\ is the dispersion relation and
$k_{c}$\ is the central momentum of the wave packet.

Now, we switch gears to the case of the present model. We note that the
geometrical phase takes a role in the dynamics and the dynamics of the wave
packet cannot be simply understood in terms of a semiclassical picture \cite%
{Bloch,Ashcroft,Kittel}. Notice that the trajectory of a wave packet is
essentially not only determined by the dispersion relation for the
field-free system but also by the geometric phase. In other words, the
dependence of the geometric phase on $k$ should be taken into account.

However, in the adiabatic limit, $\beta $ is very small. The contribution of
the geometric phase to the trajectory becomes negligible, that is,
\begin{eqnarray}
\text{Re}\gamma _{k}^{\lambda }\left( \phi \right) &\ll&\alpha _{k}^{\lambda
}\left( \phi \right).
\end{eqnarray}
The main effect of the geometric phase on the dynamics of the wave packet is
then just the modulation of the amplitude.

To verify and demonstrate the above analysis, numerical simulations are
performed to investigate the dynamics behavior. We compute the time
evolution of the wave packet by the same method as mentioned above. The
initial Gaussian wave packet has the form
\begin{equation}
\left\vert G_{k_{0}}^{N_{\mathrm{A}}}(0)\right\rangle =\frac{1}{\sqrt{\Omega
_{1}}}\sum_{j=1}^{2N}e^{-\frac{^{\alpha ^{2}}}{2}(j-N_{\mathrm{A}%
})^{2}}e^{ik_{0}j}\left\vert j\right\rangle  \label{gaussian}
\end{equation}%
with the velocity $k_{0}\in \left[ 0,2\pi \right] $, centered at the $N_{%
\mathrm{A}}$th site, where $\left\vert j\right\rangle =c_{j}^{\dagger
}\left\vert 0\right\rangle $ and $\Omega _{1}$ is a normalization factor.

For one thing, we consider the trajectories of the wave packet with
different $\beta $, as comparison to the dispersion relation. From the plots
in Fig. \ref{fig4}, we find that for small $\beta $, the trajectory accords
with the dispersion well, while as $\beta $ increases, the deviation becomes
obvious.

For another thing, we investigate the flux-controlled probability of the
wave packet. It can be rewritten in the form
\begin{equation}
\left\vert G_{k_{0}}^{N_{\mathrm{A}}}(0)\right\rangle =\sum_{k}\left(
g_{+}^{k}\left\vert \psi _{+}^{k}\right\rangle +g_{-}^{k}\left\vert \psi
_{-}^{k}\right\rangle \right) .
\end{equation}%
Here, we do not give the explicit expression of the coefficient $g_{\lambda
}^{k}$, since the following analysis is independent of $g_{\lambda }^{k}$.
Through an adiabatic evolution, we have%
\begin{equation}
\left\vert G_{k_{0}}^{N_{\mathrm{A}}}(2\pi n/\beta )\right\rangle
=\sum_{k}\left( e^{i\Omega _{+}}e^{\zeta }g_{+}^{k}\left\vert \psi
_{+}^{k}\right\rangle +e^{i\Omega _{-}}e^{-\zeta }g_{-}^{k}\left\vert \psi
_{-}^{k}\right\rangle \right) ,
\end{equation}%
where $\Omega _{\pm }$\ is a real number, and $\zeta =-\text{Im}\gamma
_{k}^{+}\left( 2\pi n\right) $. In the case of $\lambda \zeta \gg 1$, we have%
\begin{equation}
\left\vert G_{k_{0}}^{N_{\mathrm{A}}}(2\pi n/\beta )\right\rangle \approx
e^{i\Omega _{\lambda }}e^{\lambda \zeta }\sum_{k}g_{\lambda }^{k}\left\vert
\psi _{\lambda }^{k}\right\rangle .
\end{equation}%
It is shown that only the component in either upper or lower sub-band
survives. It also seems that the final state collapses to one of two
sub-bands in the context of Dirac probability. The sign of $\zeta $\ is
crucial for the direction of the collapse. We compute the evolution for two
cases with opposite flux. The result plotted in Fig. \ref{fig5} shows that
the evolved wave packet in upper or lower sub-band survives for two
different varying fluxes, which is in agreement with our prediction.

In practice, the flux control can be implemented by a pulsed flux. We
simulate this process by a Gaussian-shaped flux with the form
\begin{equation}
\phi (t)=2\sqrt{\sigma \pi }\int_{0}^{t}e^{-\sigma (t-\tau )^{2}}\text{d}t,
\label{phi_t}
\end{equation}%
which contributes $2\pi $ flux during the process. To characterize the
feature, we introduce the reduced energy%
\begin{equation}
E\left( t\right) =\frac{\sum_{k,\lambda }\varepsilon _{\lambda
}^{k}\left\vert \langle \eta _{\lambda }^{k}\left\vert G_{k_{0}}^{N_{\mathrm{%
A}}}(t)\right\rangle \right\vert ^{2}}{\sum_{k,\lambda }\left\vert \langle
\eta _{\lambda }^{k}\left\vert G_{k_{0}}^{N_{\mathrm{A}}}(t)\right\rangle
\right\vert ^{2}}.  \label{ET}
\end{equation}%
And in the adiabatic limit, we have $E\left( \infty \right) \approx
\varepsilon _{+}^{k_{0}}$ or $\varepsilon _{-}^{k_{0}}$, i.e., it converges
to a positive (negative) constant if the upper (lower) sub-wave packet
survives. Alternatively, we can replace $\varepsilon _{\lambda }^{k}$\ by $%
\lambda $\ in Eq. (\ref{ET}) to redefine $E\left( t\right) $, which leads to
$E\left( \infty \right) \approx +1$ or $-1$ even for a non-adiabatic
process. Here we take the former, because $E\left( \infty \right) $\ can
indicate the deviation between the adiabatic and non-adiabatic processes. In
Fig. \ref{fig6}, results are plotted for different values of $\sigma $,
which determine the rate of the flux change. These results clearly
demonstrate the finding of this paper that the external field can be
utilized to control the quantum state on demand via a quasi-adiabatic
process within a relatively short time.

\section{Summary}

\label{Summary} In this paper, the particle dynamics of the non-Hermitian
Rice-Mele model driven by a time-dependent external field has been
theoretically investigated. The analysis shows that the Berry phase can be a
complex number in the non-Hermitian regime. This results in the
amplification and attenuation of the amplitude of an evolved quantum state.
We found that it can have full real spectrum for any constant field, and the
Berry phase with respect to a varying field has a constant imaginary part
for an arbitrary initial state either in the upper or lower energy sub-band.
The dependence of the imaginary part of the Berry phase on the parameters,
such as lattice distortion, imaginary potential, and the direction of the
flux, was explicitly presented. Numerical simulation indicates that the
amplitude control of a wave packet can be accomplished by a quasi-adiabatic
process within a relatively short time.

\acknowledgments We acknowledge the support of the National Basic Research
Program (973 Program) of China under Grant No. 2012CB921900 and CNSF (Grant
No. 11374163).

\end{document}